\documentclass[%
 aip,
 amsmath,amssymb,
 reprint,%
]{revtex4-1}

\usepackage{graphicx}
\usepackage{xcolor}
\usepackage{dcolumn}
\usepackage{bm}
\usepackage{bbm}
\usepackage[normalem]{ulem}

\usepackage[utf8]{inputenc}
\usepackage[T1]{fontenc}
\usepackage{mathptmx}

\usepackage[colorinlistoftodos,prependcaption,textsize=tiny]{todonotes}
\newcommand{\toadd}[1]{{#1}}
\newcommand{\toremove}[1]{\ignorespaces}

\begin{document}

\title[]{
A Bayesian Inference Framework for Compression and Prediction of Quantum States
}
\author{Yannic Rath}%
\email{yannic.rath@kcl.ac.uk}
\affiliation{Department of Physics, King’s College London, Strand, London WC2R 2LS, United Kingdom}

\author{Aldo Glielmo}
\email{aglielmo@sissa.it}
\affiliation{Scuola Internazionale Superiore di Studi Avanzati (SISSA), Via Bonomea 265, 34136 Trieste, Italy}

\author{George H. Booth}
\email{george.booth@kcl.ac.uk}
\affiliation{Department of Physics, King’s College London, Strand, London WC2R 2LS, United Kingdom}%

\date{\today}

\begin{abstract}
The recently introduced Gaussian Process State (GPS) provides a highly flexible, compact and physically insightful representation of quantum many-body states based on ideas from the zoo of machine learning approaches.
In this work, we give a comprehensive description how such a state can be learned from given samples of a potentially unknown target state and show how regression approaches based on Bayesian inference can be used to compress a target state into a highly compact and accurate GPS representation.
By application of a type II maximum likelihood method based on Relevance Vector Machines (RVM), we are able to extract many-body configurations from the underlying Hilbert space which are particularly relevant for the description of the target state, as support points to define the GPS.
Together with an introduced optimization scheme for the hyperparameters of the model characterizing the weighting of modelled correlation features, this makes it possible to easily extract physical characteristics of the state such as the relative importance of particular correlation properties.
We apply the Bayesian learning scheme to the problem of modelling ground states of small Fermi-Hubbard chains and show that the found solutions represent a systematically improvable trade-off between sparsity and accuracy of the model.
Moreover, we show how the learned hyperparameters and the extracted relevant configurations, characterizing the correlation of the wavefunction, depend on the interaction strength of the Hubbard model as well as the target accuracy of the representation.
\end{abstract}

\maketitle

\section{Introduction}
The difficulty in compactly and accurately describing quantum many-body systems is a key factor limiting numerical studies of realistic physical problems found in condensed or other chemical matter.
%
This means that, in practice, numerical algorithms tackling the many-body problem need to efficiently compress the information encoded in the system to be numerically tractable.
Of particular interest is a compact description of the many-body wave function fully characterizing the physical properties of the system.
A common approach is to constrain the state to be represented by a compact functional form defining an approximate wave function ansatzes for which expectation values can still be efficiently evaluated. 
These are often built from physical characteristics which are expected (or known) to dominate the description of the target state.
Such ansatzes, including Jastrow wave function \cite{Jastrow}, Correlator Product State (CPS) \cite{Changlani2009}, coupled-cluster wave functions \cite{RevModPhys.79.291} or tensor networks \cite{Schollwoeck2011,Orus2014,Sandvik2007} are usually designed to explicitly model selected correlations (such as those based around low-rank, locality or low-entanglement) of the target system which makes them highly successful for specific applications.

Recent approaches to describe many-body wave functions have also used the idea of deriving their functional forms from the field of machine learning (ML), with the aim of avoiding high prior constraints on the emergent physics.
Such ML-based approaches often represent a direct conceptual counterpart to other numerical techniques; instead of incorporating as much physical understanding as possible in order to reduce the complexity of the problem, algorithms are introduced to automatically find optimal approximations for the given problem based on data or some quality metrics in a non-parametric fashion.
This includes the representation of states based on artificial neural networks, typically denoted as Neural Quantum States (NQS) \cite{Carleo2017, Chen2018, Levine2019, Jia2019, Pastori2019, Carrasquilla2020} which have successfully modelled the ground states of lattice \cite{Nomura2017, Cai2018, Liang2018, Sharir2020, Yang2020, Westerhout2020, Zen2020, Szabo2020, HibatAllah2020, Kochkov2018, Roth2020} and ab-initio \cite{Schuett2019, Choo2020, Pfau2019, Hermann2019} many-body systems, as well as been used in quantum state tomography \cite{Torlai2018}, and a description of dynamical properties of many-body systems \cite{Hendry2019, Zhang2020}.
Complementary to the representations based on different neural network architectures, an additional class of many-body wave functions motivated by machine learning with kernel methods, in particular Gaussian process regression, has been introduced recently.
These states, named Gaussian Process States (GPS) \cite{Glielmo2020}, are explicitly data-driven, and have a number of appealing properties. In particular, it is easy to incorporate anticipated physical intuition into the method to improve its efficiency, while the final description remains automatically learned and \toadd{can in practice be constructed to be systematically improvable to exactness.}
The GPS have already shown promise in the description of many-body ground states which we substantiate and elaborate on in this work.
In particular, we describe how the particular form of the GPS makes it well suited for the application of regression techniques based on Bayesian inference to obtain highly accurate and compact many-body representations from given wavefunction data which can be used to compute physical quantities of interest.

%
In this work, we benchmark the accuracy of the Bayesian procedure to efficiently compress a given many-body wave function to GPS form, by learning ground states of one-dimensional Fermi-Hubbard models. This defines a clear series of correlated and controllable test systems with known exact comparison results, defined by the Hamiltonian
\begin{equation}
	\hat{H} = -t \sum_{\langle i,j\rangle,\sigma} {\hat c}_{i,\sigma}^{\dagger} {\hat c}_{j,\sigma} + U\sum_{i} \hat{n}_{i \uparrow} \hat{n}_{i \downarrow},
\end{equation}
where Fermionic creation and annihilation operators $c^\dagger$ and $c$ are associated with discrete sites of a one-dimensional lattice, labelled by indices $i$ and $j$, and with spins labelled by $\sigma$.
The model, characterized by a hopping parameter $t$ and the interaction strength $U$, provides a generic prototype for the description of strongly correlated electrons in the context of condensed matter physics and phenomena emerging in such systems.
Numerical studies of this system face the many-body problem as they must find a compact (polynomial) description of the system while still being able to correctly describe the emergent correlated effects.
We thus aim to represent the ground state as a compact map in the form of a GPS which associates a wave function amplitude to any many-body configuration of the underlying Hilbert space.

The presented work expands on the GPS framework introduced in Ref.~\onlinecite{Glielmo2020}, which is then extended to gain further insight into this novel Bayesian approach to wave function representation. We show that this framework also allows us to explore the potential to extract interpretable physical information from such a state, including correlation length scales, and showcases the potential of the GPS representation to tackle the many-body problem.

\section{Gaussian process states}
\subsection{The GPS representation}
The GPS models the logarithm of the wave function amplitude as the mean of a Gaussian process regression estimate \cite{Rasmussen_book}.  This defines the amplitude for a many-body configuration $\mathbf{x}$ as
\begin{equation}
    \psi(\mathbf{x}) = e^{\sum_b w_b k(\mathbf{x}, \mathbf{x}_b')},
    \label{eq:GPS}
\end{equation}
where the sum in the exponential is taken with respect to a set of `basis' configurations $\mathbf{x}_b'$, which define the support points of the model. This model is therefore explicitly data-driven, with each basis configuration in the data set also having an associated weight $w_b$. The other quantity required is the kernel function $k(\mathbf{x}, \mathbf{x}_b')$, which compares the test configuration $\mathbf{x}$ with the basis configurations.
Since the ground state of Hubbard chains considered in this work can be made to be entirely positive in the site basis if appropriate boundary conditions are chosen, we do not require to learn sign structures of the wave function. Learning the sign structure of the wave function amplitudes -- a challenging task for both the GPS and NQS representations \cite{Westerhout2020, Szabo2020} -- is a topic we defer to a later study after consideration here of the ability to learn the exponential space of positive-definite amplitudes in this form. We can therefore restrict the weights of the model to be real in this instance.

The estimator $\sum_b w_b k(\mathbf{x}, \mathbf{x}_b')$ can be considered a linear regression model in some (typically high dimensional) space of features of the many-body configurations.
%
%
However, the feature map, transforming many-body configurations into this feature space, is only defined implicitly through the kernel function $k(\mathbf{x}, \mathbf{x}_b')$ which corresponds to a scalar product between the feature vectors associated with $\mathbf{x}$ and $\mathbf{x}_b'$.
An important property of this representation is that the transformation of the configurations into the feature space does not need to be evaluated explicitly.
This means that, in the context of many-body wave functions, we can define kernel functions which represent an exponentially scaling number of correlation features emerging between local degrees of freedom of a system, while still maintaining a functional form which can be evaluated efficiently for any many-body configuration.
This ensures that if an appropriate kernel function is chosen, a GPS can represent any wave function to arbitrary accuracy, making it a representation which is not limited in its expressibility.
Such a general exact representation of arbitrary states usually requires a number of basis configurations which grows as the dimensionality of the underlying Hilbert space.
Nonetheless, we aim to obtain compact but still highly accurate GPS representations by efficiently modelling the underlying physical structure of relevant states, mirroring the great success of function approximators based on kernel functions used in different machine learning applications, e.g. with Gaussian process regression \cite{Glielmo2017}, kernel ridge regression \cite{Vu2015} or support vector machines \cite{Ponte2017}.
The main task discussed in the following is to learn a GPS representation of a given physical state which is as compact (in terms of a small number of basis configurations) and accurate as possible.

\subsection{The kernel function}
A central element to the success of the method is the choice of the kernel function, defining the feature space in which the linear regression of the log-wave function amplitudes takes place.
The features which are described with the ansatz are therefore defined before the actual representation is learned from the given data.
%
This is
\toadd{conceptually} different from \toadd{how typical} representations based on artificial neural networks \toadd{are used in practice.}
\toadd{Although the features learned for such states could also be designed by choosing particular input representations or specific network architectures,} 
\toadd{usually their} features are in principle learned for a specific state via a Monte Carlo sampling and variational optimization of the chosen network architecture.
However, being able to define the kernel function allows for a high level of \toadd{direct} control over the underlying physics which is modelled and they can be explicitly tailored to enforce desired physical constraints on the wave function.
For example, as shown in Ref.~\onlinecite{Glielmo2020}, many-body states such as the W-state or the Laughlin state can be represented very efficiently as a GPS using only a single basis configuration by introducing a specifically designed kernel function.
Moreover, it was found that general kernel functions (not tailored for a specific state) can learn the ground state wave functions efficiently.
This can be attributed to the fact that the kernel transforms the configurational input into a very high (or even infinite) dimensional feature space, where there always exists a solution which is of linear form. Furthermore, similar to an NQS representation, it is possible to include a degree of `kernel learning', in order to optimize hyperparameters of the kernel, resulting in a weighting of some more relevant features over others, with their values giving insight into the nature of the state.
%

%
In the context of many-body wave functions, the key challenge is that of finding regularities between the wave function amplitude of a given many-body configuration and the local occupancies of its sites.
If there are correlations between amplitude and occupancy patterns, then we expect to be able to exploit this information and obtain a compressed representation of the state.
An example of the use of just such a correlation features can be found in the Gutzwiller representation \cite{Gutzwiller1963} (as one example) which explicitly models the exponential suppression (or enhancement) of the wave function amplitude depending on the number of doubly occupied sites in the configuration.
With the GPS we are not limited to the extraction of such single-site features, and we can in principle represent and model correlations between an arbitrary number and range of lattice sites.
In this way, the kernel can represent a sum over exponentially many features, with the exponential form of the final wave function form of Eq.~(\ref{eq:GPS}) ensuring overall product separability of these features and a resulting state which preserves size extensivity of appropriate properties (such as the total energy).

With this aim, in this work we choose to use a squared exponential kernel within a GPS framework.
The base kernel is defined as
\begin{equation}
    k(\mathbf{x}, \mathbf{x}') = e^{- h(\mathbf{x}, \mathbf{x}')/\theta},
\end{equation}
where $h(\mathbf{x}, \mathbf{x}')$ is the ``Hamming distance'' \cite{Hamming1950} between the many-body configurations $\mathbf{x}$ and $\mathbf{x}'$. Furthermore, $\theta$ is an adjustable hyperparameter controlling the weighting of the modelled features at different ranks.
For the Fermi-Hubbard model, we denote the local occupancy of configuration $\mathbf{x}$ at site $i$ as $x_i$, which can take one of four possibilities, \mbox{$\{\cdot, \uparrow, \downarrow, \uparrow \downarrow\}$}. The Hamming distance between two configurations can then be constructed as
\begin{equation}
    h(\mathbf{x}, \mathbf{x}') = \sum_i (1 - \delta_{x_i, x'_i}),
\end{equation}
where the sum ranges over all sites of the lattice and the delta function $\delta_{x_i, x'_i}$ is equal to one if the local occupancies at site $i$ of configurations $\mathbf{x}$ and $\mathbf{x}'$ are the same and to zero otherwise.

Such a kernel gives rise to a feature space which is a linear combination of all possible occupancy configurations across arbitrary many sites (derived in more detail in Ref.~\onlinecite{Glielmo2020}) which guarantees that an exact representation can always be obtained in the limit of a complete set of basis configurations.
This property can be easily seen by considering the Taylor expansion of the kernel function
\begin{equation}
    k(\mathbf{x}, \mathbf{x}') = 1 - \frac{\sum_i (1 - \delta_{x_i, x'_i})}{\theta} + \frac{\sum_{i,j} (1 - \delta_{x_i, x'_i})(1 - \delta_{x_j, x'_j})}{2 \theta^2} - \ldots.
    \label{eq:squared_exp_kernel}
\end{equation}
In the second term only single-site correlation features in the two configurations are represented so that these terms alone would give rise to a Gutzwiller-type parametrization of single-site features.
The third term also contains terms of the form $\delta_{x_i, x'_i} \delta_{x_j, x'_j}$ thus also capturing two-site correlation features between different sites and similarly higher order correlation features are contained in the following terms.
The $n$-th order of correlations between different sites captured in the kernel is suppressed by a factor of $\theta^n$ so that the hyperparameter $\theta$ effectively controls the weighting of higher order correlation features with respect to the lower order features.

In addition to the weighting of the correlation order in the kernel function, we can introduce another hyperparameter controlling the relative weighting of correlation features based on the distance between sites, since we expect short range correlations to be dominant over the long range correlations.
We parametrize this effect through an explicit polynomial suppression of long range features in the Hamming distance as
\begin{equation}
    h(\mathbf{x}, \mathbf{x}') = \sum_i \frac{(1 - \delta_{x_i, x'_i})}{\vert r_i - r_0 \vert^\gamma},
    \label{eq:modified_HD}
\end{equation}
where $\vert r_i - r_0 \vert$ represents the distance of the site with index $i$ from a chosen reference site with index zero. 
\toadd{For positive values of $\gamma$, this construction explicitly introduces a stronger weighting of features resulting from the occupancy patterns around the chosen reference site.
However, if the basis configurations of the ansatz are chosen from the set of all possible configurations (including translationally equivalent ones), then the chosen reference site will not bias or affect the accuracy of the ansatz, since the model can simply shift the data set of basis configurations to compensate for any change in reference choice. Therefore, for translationally symmetric systems and a free choice for the model to pick desired basis configurations, the model is independent of the arbitrary choice of reference site.}

\toadd{The choice of basis configurations for the model will thus most sensitively depend on their occupancies around this arbitrary reference site, where the}
hyperparameter $\gamma$ explicitly controls the rate of decay in the relative importance of long range compared to short range features from this reference site. It should be stressed that the adjustment of these hyperparameters ($\gamma$ and $\theta$) does not restrict the dimensionality of the feature space, or bias the exact limit of the ansatze in the limit of complete and exact training data and basis size. However, the convergence to this exact limit with increasing data sets can be improved if the hyperparameters are set such that they preferentially weight the fitting of these more important features. This is because decreasing these hyperparameters leads to a more complex model, which requires a larger basis set to appropriately fit to a desired accuracy. This will be elaborated on in Sec.~\ref{sec:Hyperparams}.

Using this modified, distance-regularized Hamming distance defined in Eq.~(\ref{eq:modified_HD}) within the squared exponential form for the kernel (taking care with the limit of $r_i \to r_0$), we obtain the final form of our kernel, as
\begin{equation}
    k(\mathbf{x}, \mathbf{x}') = \delta_{x_0, x'_0} \exp{\left(- \sum_{i > 0} \frac{1 - \delta_{x_i, x'_i}}{\theta \, \vert r_i - r_0 \vert^\gamma} \right)}.
    \label{eq:final_kernel}
\end{equation}
An important property of this kernel function is that the basis configurations $\mathbf{x}'$ do not necessarily need to correspond to physical configurations associated with the target Hamiltonian.
By building basis configurations associated with a lattice smaller than the target system (and restricting the sum in the exponential to site indices of the smaller lattice), we can explicitly restrict the range of the modelled correlation features as it is typically done in the framework of Correlator Product States (CPS) \cite{Changlani2009}.
However, contrary to a CPS ansatz, the kernel in Eq.~\eqref{eq:final_kernel} allows for modelling the exponentially growing number of features across the considered range in a polynomially compact form.
This is due to the fact that the kernel, which can be evaluated in polynomial time, implicitly defines the map into the space of described features independently of the number of basis configurations used in the GPS.

%
Finally, physical symmetries of the system can easily be incorporated into the GPS representation by explicitly encoding them into the kernel function.
A direct way to do this is to include a sum over all symmetry operations in the kernel definition, giving the symmetrized kernel
\begin{equation}
    k_{sym}(\mathbf{x}, \mathbf{x}') = \sum_{\mathcal{S}} k(\mathcal{S}[\mathbf{x}], \mathbf{x}'),
\end{equation}
where the operations $\mathcal{S}$ represent the symmetry operations we want to respect and $\mathcal{S}[\mathbf{x}]$ denotes the transformed configuration which is obtained after applying $\mathcal{S}$ to the configuration $\mathbf{x}$.
In this work, we always include translational symmetry of the lattice into the kernel through this construction, ensuring that the GPS state preserves translational symmetry at all times.
We note that this kernel is then equivalent to the `complete' kernel introduced in Ref.~\onlinecite{Glielmo2020}, as the $p \rightarrow \infty$ limit (Eq.~7 of that work), modelling all correlations, with a specific choice ($\vert r_i - r_0 \vert^{-1}$) for the distance weighting of the features. 
%

\section{Bayesian compression of quantum states}
\subsection{Bayesian learning with GPS}
In \toadd{this}
section we address the problem of finding a suitable GPS representation from a given set of wave function data corresponding to a target state.
This is essentially a task of supervised learning: given a finite set of input-output pairs we want to infer a good general model describing the underlying relationship between the inputs and the outputs.
Specifically here, we want to learn a good GPS representation of the ground state of the Hubbard model from a given set of wave function amplitudes, e.g. generated by exact diagonalization. 
We tackle this problem by applying well known machine learning techniques based on Bayesian inference \cite{Tipping_2004, Rasmussen_book} which have successfully been used for various different regression and classification tasks.

The essential paradigm of such learning schemes is to explicitly model a probability distribution over all possible models which is statistically inferred from the given data.
Applying such Bayesian learning schemes we therefore do not only obtain a GPS which describes the given amplitudes well (e.g. by adopting the most probable or the mean of all models) but we also gain insight about underlying statistical properties such as the uncertainty for new predictions associated with a chosen model.
This additional interpretability of the results is a key advantage over non-statistical regressions schemes such as a simple minimization of the squared errors on the training data.

In order to model the GPS as defined in Eq.~\eqref{eq:GPS}, we require a set of basis configurations (which is hopefully optimally compact) along with associated weights, $\{(\mathbf{x}_b', w_b)\}$, as well as the kernel function which is in this case parameterized by the hyperparameters $\theta$ and $\gamma$.
The objective is therefore to describe a probability distribution over all possible choices of basis configurations, associated weights and kernel hyperparameters defining the model, given (some or all) `training' wave function amplitudes $\{\psi_T(\mathbf{x})\}$ of a target state $\psi_T$. The aim is for this GPS model to use the training data in order to faithfully represent this target state in a compact form. In general, this target state will be approximate, or even unknown. However, in this work we will consider an exact $\psi_T$, to allow for exploration of optimal algorithms, and unambiguous conclusions into the underlying efficiency and accuracy of the model. The development of practical algorithms for unknown states based on this approach is explored in Ref.~\onlinecite{Glielmo2020}, and will be the subject of future work.

Instead of fitting directly on the set of training wave function amplitudes $\{\psi_T(\mathbf{x})\}$, we apply the Bayesian regression scheme to the logarithm of the amplitudes, $\phi(\mathbf{x}) = \ln{\psi_T(\mathbf{x})}$, with the linear model $\mathcal{M} \equiv \sum_b w_b k(\mathbf{x}, \mathbf{x}_b')$.
The central modelling assumption is that the linear model $\mathcal{M}$ (which depends on the basis configurations, associated weights and kernel hyperparameters) approximates the log-wave function amplitudes well, and that the probability distribution for the error in each log amplitude follows an independent normal distribution with zero mean and a variance parameterized by $\sigma^2_{\mathbf{x}}$.
This assumption results in a multivariate normal distribution with diagonal covariance matrix for the likelihood of the training data, $p(\bm{\phi} \vert \mathcal{M}, \bm{\sigma^2})$. This is the probability of observing the given training log-wave function values (here denoted by the vector $\bm{\phi}$) from a particular set of model variables and associated variances (denoted by the vector $\bm{{\sigma^2}}$).
In typical Bayesian regression, one would use a single noise parameter $\sigma^2$, describing the variance of the likelihood when it is expected that the distribution of errors for this quantity is independent of the specific configuration.
In our case however, we are modelling the logarithm of the wave function and are rather interested in a fixed variance for the description of the {\em actual} amplitudes and not necessarily for their logarithms. We can achieve this by choosing $\sigma^2_{\mathbf{x}}$ in order to unbias the action of the logarithm on the variances, and to instead fix an (approximately) constant variance in the likelihood of modelling the true wave function amplitudes of $\psi_T(\mathbf{x})$.

To do this, we note that the chosen assumptions result in a log-normal distribution for the likelihood of the actual wave function amplitudes, with variance
\begin{equation}
    \text{Var}(\psi_T(\mathbf{x})) = \exp(\sigma_{\mathbf{x}}^2-1) \langle \psi_T(\mathbf{x}) \rangle^2,
\end{equation}
where $\langle \psi_T(\mathbf{x}) \rangle$ represents the mean of the likelihood for the wave function amplitude from the model.
This means that the variance of the distribution generally increases with increasing mean if we use a fixed value of $\sigma_{\mathbf{x}}^2$ for all configurations.
In order to obtain a fixed variance, we can solve the equation for $\sigma_{\mathbf{x}}^2$ and introduce a single parameter $\tilde{\sigma}^2$ which defines the target variance we want to obtain for the likelihood of the actual wave function amplitudes.
This gives the rescaled noise parameter for the variance of the likelihood, as
\begin{equation}
    \sigma_{\mathbf{x}}^2 = \ln{ \left(1 + \frac{\tilde{\sigma}^2}{\langle \psi_T(\mathbf{x}) \rangle^2} \right)}.
\end{equation}
Under the assumption that the model fit describes the training data reasonably well, $\langle \psi_T(\textbf{x}) \rangle$ can be approximated using the value of the training amplitude, giving the final rescaled noise as $\sigma_{\mathbf{x}}^2 = \ln{(1 + \frac{\tilde{\sigma}^2}{\psi_T(\mathbf{x})^2})}$.
The target variance of the likelihood for the actual wave function amplitudes, $\tilde{\sigma}^2$, can then be used as an input parameter defining the desired accuracy which should be achieved by the fit.
A larger value of $\tilde{\sigma}^2$ corresponds to a larger variance of the likelihood so that the data might be described by a less accurate model and vice versa. This therefore represents the key value determining the accuracy of the final model.

\subsection{Bayesian optimization of the model weights} \label{sec:weight_opt}
Using the modelling assumptions above, we can now address the problem of finding the \textit{posterior} distribution, $p(\mathcal{M} \vert \bm{\phi}, \tilde{\sigma}^2)$, describing the probability distribution over all possible models, given the training data and a target accuracy defined by $\tilde{\sigma}^2$. The mean of this distribution will define the final GPS model parameters required for the wave function in Eq.~\eqref{eq:GPS}.
In order to obtain the posterior distribution, we apply Bayes' theorem, giving the posterior as
\begin{equation}
    p(\mathcal{M} \vert \bm{\phi}, \tilde{\sigma}^2) = \frac{p(\bm{\phi} \vert \mathcal{M}, \tilde{\sigma}^2) \times p(\mathcal{M} \vert \tilde{\sigma}^2)}{\int d \mathcal{M} \, p(\bm{\phi} \vert \mathcal{M}, \tilde{\sigma}^2) \times p(\mathcal{M} \vert \tilde{\sigma}^2)},
\end{equation}
with likelihood $p(\bm{\phi} \vert \mathcal{M}, \tilde{\sigma}^2)$ and where $p(\mathcal{M} \vert \tilde{\sigma}^2)$ denotes the \textit{prior} which is our assumed probability distribution over the model variables in the absence of any data.
The denominator of the expression, known as the \textit{marginal likelihood}, normalizes the probability distribution.

As a first application of Bayesian inference, we can find the posterior distribution for the weights of the model (written as a vector $\bm{w}$) for a fixed basis set and fixed hyperparameters $\theta$ and $\gamma$.
Without including any prior belief about the distribution of the weights (i.e. using a uniform prior, $p(\bm{w})$), the posterior follows a normal distribution, with mean equivalent to a direct minimization of the squared errors weighted by the associated values of $\sigma_{\mathbf{x}}^2$ over the training set.
This minimization can be performed in closed form, with the most probable weights then given by
\begin{equation}
    \bm{w}_{MP} = \mathop{\mathrm{argmin}}_{\bm{w}} \sum_{\mathbf{x}} \frac{ \vert\phi(\mathbf{x}) - \mathcal{M}(\mathbf{x}) \vert ^2}{\sigma_{\mathbf{x}}^2} = (K^T S^2 K)^{-1} K^{T} S^2 \bm{\phi},
\end{equation}
where $K$ is the kernel matrix evaluated between all basis and training configurations, and $S^2$ is a diagonal matrix with elements $\frac{1}{\sigma_{\mathbf{x}}^2}$.

In addition to the problem of potentially obtaining (numerically) singular matrices $K^T S^2 K$, it is also known that such least squares solutions are prone to overfitting the training data, so that the obtained model does not necessarily generalize well for the prediction of configurations not contained in the training data set.
These problems can both be addressed by introducing an appropriate prior distribution to regularize the optimal weights.
We employ the common choice of a Gaussian prior with zero mean and a variance $\alpha^{-1}$. This form ensures that the posterior distribution remains analytically tractable, and introduces a bias towards learned models with weights distributed around zero.
With this choice, the posterior distribution for the weights is also Gaussian, with a mean for the distribution given by
\begin{equation}
    \bm{\mu} = (K^T S^2 K + \alpha \mathbbm{1})^{-1} K^{T} S^2 \bm{\phi},
    \label{eq:ridge_estimator}
\end{equation}
and covariance matrix
\begin{equation}
    \Sigma = (K^T S^2 K + \alpha \mathbbm{1})^{-1}.
    \label{eq:posterior_covariance}
\end{equation}
The mean weights, which are adopted as the weights for the model, are equivalent to the ones obtained with a ``regularized'' least squares fit where the sum over the squared errors is optimized subject to a regularization constraint for the sum of all squared weights - a procedure known as ridge regression \cite{Hoerl_1970} or Tikhonov regularization.
In order to allow for models satisfying the imposed zero-mean prior distribution of the weights, we shift our training data before the fit, so that the mean over all training amplitudes $\phi(\mathbf{x})$ is zero.
This only affects the normalization of the final GPS wave function and can either be reverted by shifting back the log-amplitudes when predictions are made, or simply ignoring this shift if the final quantities are independent of the overall wave function norm.

\begin{figure}
    \centering
    \includegraphics[width=\columnwidth]{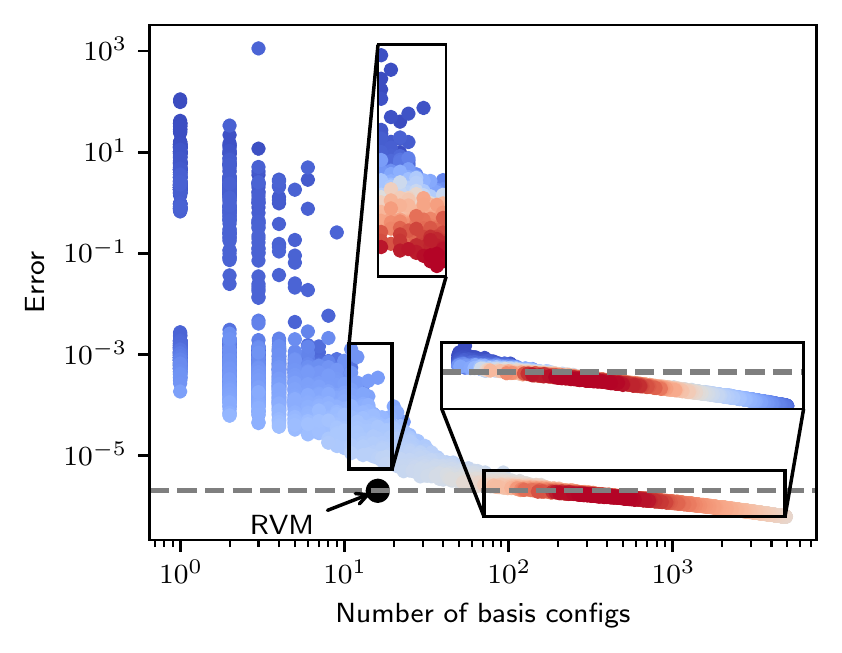}
    \caption{Scatterplot of the mean squared error obtained for a GPS trained on the exact ground state of an eight site Hubbard chain at $U/t =8$ against the number of randomly chosen basis configurations in the model. The GPS was trained by inferring the weights according to Eq.~\eqref{eq:ridge_estimator} for a total of $10,000$ different randomly sampled basis sets (each corresponding to a point) using fixed values $\gamma = 1$, $\theta = 10$, $\alpha=1$ and $\tilde{\sigma}^2=10$. The coloring of scatter points corresponds to the log marginal likelihood with larger values marked in red and smaller values shown in blue. The colorscale of the marginal likelihood is rescaled for the two insets showing particular portions of the full scatterplot. Also shown is the result obtained by application of the relevance vector machine to find the optimal basis set configurations via a maximum marginal likelihood scheme (Sec.~\ref{sec:RVM}) with the same value of $\tilde{\sigma}^2$ and kernel hyperparameters.}
    \label{fig:least_squares_scatterplot}
\end{figure}

Within this Bayesian approach, we are therefore able to obtain the optimal weights for a fit to the training data of a target state, given some basis configurations, kernel hyperparameters, as well as variances for the likelihood of the wave function amplitudes, $\tilde{\sigma}^2$, and prior for the weights, characterized by $\alpha$.
An illustration of the quality of this resulting model corresponding to a regularized least squares fit, is visualized in Fig.~\ref{fig:least_squares_scatterplot}. This shows the error in the overall modelled wave function amplitudes compared to the exact target ground state wave function, for the highly correlated $U/t=8$ one-dimensional Hubbard model with eight sites, as the number of basis configurations defining the model increases. Each point corresponds to a model where the desired number of basis configurations are chosen entirely at random from all possible configurations of the underlying configurational Hilbert space. The weights are then obtained by the mean of the posterior distribution as defined in Eq.~\eqref{eq:ridge_estimator}, with the training data consisting of the entire set of symmetrically inequivalent amplitudes of the target state.

The first aspect which can be seen from the figure is that the quality of the fit (represented by the mean squared error obtained with the GPS across the Hilbert space) varies significantly for different random basis sets, giving mean squared errors ranging between $\approx 10^3$ and $\approx 6 \times 10^{-7}$.
As expected, we achieve the smallest errors for the fit of the target state in the limit of large basis sets, which underlines a very general intuition about the GPS: With a larger number of basis configurations defining the model, the better the expressibility of the model, allowing for a higher accuracy in the fit to the target state.
However, we are usually not interested in the limit where almost all of the configurations from the Hilbert space are used as basis configurations for our ansatz but aim to achieve a more compact representation of the target state.
Whereas the error barely fluctuates for larger basis sets where a significant fraction of all of possible configurations are included in the chosen basis set, the mean squared errors fluctuate heavily for small basis sets corresponding to compact models.
For example, a model defined with a single randomly chosen basis configuration achieves an error ranging between $\approx 2 \times 10^{-4}$ and $\approx{100}$, indicating that not all basis set choices are equally well suited to achieve a compact but still accurate representation of the target state.
It is clear (as expected) that an appropriate selection of basis configurations for a compact model is of paramount importance to its success.
Learning a GPS from given data therefore also involves finding a suitable set of basis configurations (and potentially also kernel hyperparameters), which is the topic we turn to in the next section.

\subsection{Type II maximum likelihood}
Ideally, we would infer the full model, including choice of basis configurations and kernel hyperparameters, in the same Bayesian approach as the weights, i.e. by defining a sensible prior for the full model and finding the model which maximizes the posterior distribution $p(\mathcal{M} \vert \bm{\phi}, \tilde{\sigma}^2)$, given the training data.
However, this approach would not be tractable in practice, requiring the introduction of additional modelling assumptions.
Separating the known posterior distribution for the weights, the full posterior for the model can be written as
\begin{equation}
    p(\mathcal{M} \vert \bm{\phi}, \tilde{\sigma}^2) = p(\bm{w} \vert \{\mathbf{x}_b'\}, \theta, \gamma, \bm{\phi}, \tilde{\sigma}^2) \times p(\{\mathbf{x}_b'\}, \theta, \gamma \vert \bm{\phi}, \tilde{\sigma}^2),
\end{equation}
where we introduce the posterior for the basis set and the kernel hyperparameters as $p(\{\mathbf{x}_b\}, \theta, \gamma \vert \bm{\phi}, \tilde{\sigma}^2)$.
Instead of maximising the full posterior of the overall model, $p(\mathcal{M} \vert \bm{\phi}, \tilde{\sigma}^2)$, we apply the common type II maximum likelihood procedure. This means that we find the basis set and hyper parameters which maximise $p(\{\mathbf{x}_b\}, \theta, \gamma \vert \bm{\phi}, \tilde{\sigma}^2)$, which is independent of the weights, and then use the found maximum of this distribution for inference of the resulting weights.
If we assume uniform priors for the basis sets and kernel hyperparameters, then the posterior $p(\{\mathbf{x}_b\}, \theta, \gamma \vert \bm{\phi}, \tilde{\sigma}^2)$ is proportional to the marginal likelihood which appears in the application of Bayes' theorem for the inference of the weights,
\begin{equation}
    p(\{\mathbf{x}_b'\}, \theta, \gamma \vert \bm{\phi}, \tilde{\sigma}^2) \propto \underbrace{\int d \bm{w} \, p(\bm{\phi} \vert \mathcal{M}, \tilde{\sigma}^2) \times p(\bm{w} \vert \alpha)}_{p_{ML}}.
\end{equation}
We therefore aim to find an optimal and compact basis set and kernel hyperparameters by maximising the marginal likelihood $p_{ML} = p(\bm{\phi} \vert \{\mathbf{x}_b'\}, \theta, \gamma, \tilde{\sigma}^2, \alpha)$ for the weights.
Assuming a Gaussian likelihood and prior for $\bm{w}$ as done in Sec.~\ref{sec:weight_opt}, the marginal likelihood can be obtained in closed form for a chosen basis set.
Its logarithm is then given by \cite{Tipping2003}
\begin{equation}
   \ln{(p_{ML})} =  -\frac{1}{2} (N \ln{(2 \pi)} + \ln{(\det{(C)})} + \bm{\phi}^T C^{-1} \bm{\phi}),
    \label{eq:log_ml_slow}
\end{equation}
where $N$ denotes the number of training amplitudes and where
\begin{equation}
    C = {(S^{2})}^{-1} + K (\alpha \mathbbm{1})^{-1} K^T.
\end{equation}
An alternative form of the log marginal likelihood which can in practice be faster to evaluate is given by \cite{fletcher2010relevance}
\begin{multline}
   \ln{(p_{ML})} =  \frac{1}{2} (N_b \ln{(\alpha)} - \sum_{\{\mathbf{x}\}} \ln{(2 \pi \sigma_{\mathbf{x}}^2)} \\ + \ln{(\det{(\Sigma)})} - \bm{\phi}^T S^2 \bm{\phi} + \bm{\mu}^T \Sigma^{-1} \bm{\mu}).
    \label{eq:log_ml_fast}
\end{multline}
Here $N_b$ denotes the number of basis configurations, the set $\{\mathbf{x}\}$ represents the set of all training configurations with associated log-amplitudes $\bm{\phi}$, and $\bm{\mu}$ and $\Sigma$ are the mean and covariance matrix as given in Eq.~\eqref{eq:ridge_estimator} and Eq.~\eqref{eq:posterior_covariance}.

The ability of the marginal likelihood to distinguish the quality of different basis sets can also be seen in Fig.~\ref{fig:least_squares_scatterplot} for the randomly selected basis sets of varying sizes.
The marginal likelihood obtained for each basis set is indicated by the coloring of the different scatter points with warmer colors representing a larger marginal likelihood.
As can be seen in the left inset, in the regime of smaller basis sets, the computed value of the marginal likelihood is typically larger for those basis sets which reach a higher accuracy compared to poorer quality basis sets of the same size. 
The marginal likelihood thus provides a good figure of merit to gauge the quality of a basis set of a given size.
Overall, the computed value of the marginal likelihood for small basis sets increases with higher accuracy of the model, reaching a maximum at $\approx 360$ configurations.
For basis sets significantly larger than this, the marginal likelihood then decreases.
This shows that the marginal likelihood is not simply inversely corresponding to the error obtained for the fit, but also attempts to maximise the sparsity of the model, i.e. to minimize the number of chosen basis configurations.
This well known property of the marginal likelihood under our chosen modelling assumptions \cite{Tipping_2004}, makes statistical inference of the weights together with optimization of the marginal likelihood the ideal candidate for compressing a target state into a compact and representative GPS. If a higher accuracy (and correspondingly larger basis set) is desired, then this can be controlled with the ${\tilde \sigma}$ parameter. Decreasing this value will specify that we desire a more accurate fitting to the training data (reducing the variance in the likelihood of the model reproducing the data). Consequently, the maximization of the marginal likelihood will result in the selection of larger basis sets with higher accuracy in the fitting of the state. Consideration of the effect of varying ${\tilde \sigma}$ will be considered in more detail in Secs.~\ref{sec:Hyperparams} and \ref{sec:FullOpt}.

\subsection{The Relevance Vector Machine for basis set selection} \label{sec:RVM}

Finding the maximum of the marginal likelihood directly with respect to all possible basis sets is computationally intractable due to the combinatorial number of possible basis sets.
However, it is possible to identify optimally relevant basis configurations from the set of all possible configurations by application of the \textit{Relevance Vector Machine} (RVM) \cite{Tipping2000}.
Within the RVM, the prior of the weights is not modelled by a single variance parameter for all weights, $\alpha$, but rather is allowed to vary between configurations. This modifies the prior distribution for each weight as a normal distribution with zero mean, but an individual variance of $1/\alpha_{\mathbf{x}'}$ which now depends on the configuration, $\mathbf{x}'$.
The full prior for the weights is therefore given by a multivariate normal distribution with zero mean and covariance matrix $A^{-1}$, where $A$ is a diagonal matrix with elements $\alpha_{\mathbf{x}'}$.
This results in the same posterior distribution for the weights, with mean, covariance matrix and log-marginal likelihood as defined in Eq.~\eqref{eq:ridge_estimator}, Eq.~\eqref{eq:posterior_covariance} and Eqs.\eqref{eq:log_ml_slow} and \eqref{eq:log_ml_fast}, but where the matrix $\alpha \mathbbm{1}$ is replaced by the diagonal matrix $A$.
The type II maximum likelihood procedure then involves finding the maximum of the marginal likelihood with respect to the parameters $\alpha_{\mathbf{x}'}$ for all the potential basis configurations $\{\mathbf{x}'\}$.
If the optimal parameters $\alpha_{\mathbf{x}'}$ for a basis configuration $\mathbf{x}'$ goes to infinity, the prior for the weight associated with this configuration becomes a delta function located at zero.
This implies that the optimal weight for this candidate configuration tends to zero, so that this basis configuration is not relevant for the model and it can be removed.
The final basis set obtained by the RVM from a set of (potentially all) candidate configurations are then those configurations with finite optimised parameter $\alpha_{\mathbf{x}'}$.
In practice it can be observed that indeed many of the $\alpha_{\mathbf{x}'}$ parameters go to infinity during the optimisation of the marginal likelihood, which is another manifestation of the observation that a large marginal likelihood represents an optimization of not just the accuracy of the model, but also the sparsity of the chosen configurational basis set, as dictated by the sparse form of the prior on the weights.

In this work, we apply the fast marginal likelihood optimization algorithm presented in Ref.~\onlinecite{Tipping2003}.
This starts by setting all but one of the $\alpha_{\mathbf{x}'}$ to infinity before the marginal likelihood is optimized iteratively by updating a single value $\alpha_{\mathbf{x}'}$ at each iteration step until the algorithm converges.
Due to closed analytic forms for the marginal likelihood, updates for the value of a selected $\alpha_{\mathbf{x}'}$ at each step can be found analytically, while it is also possible to identify the configuration which will give the largest improvement to the overall marginal likelihood under the update of its $\alpha_{\mathbf{x}'}$ value. This allows for a very efficient algorithm to find the optimal basis set.
If $\alpha_{\mathbf{x}'}$ is updated from infinity to a finite value then this configuration is added to the set of active basis configurations and if the value is updated from a finite value to infinity then the configuration is removed from the active basis set.
A key advantage of this fast iterative optimization scheme is that the set of active basis configurations is typically kept small at all times throughout the optimization. 
This reduces the computational cost of the algorithm because it involves only the inversion of a $N_{active} \times N_{active}$ matrix at each iterative step where $N_{active}$ is the number of active basis configurations (i.e. basis configurations with finite $\alpha_{\mathbf{x}'}$).

The success of the RVM to obtain a sparse set of relevant basis configurations is exemplified by returning to the scatterplot of Fig.~\ref{fig:least_squares_scatterplot}. This also includes the result obtained with the RVM using the same kernel and accuracy hyperparameters as for the values obtained with randomly selected basis sets.
The RVM, selecting the most relevant of all possible basis configurations, gives a GPS with a mean squared error on the target state of $\approx 2 \times 10^{-6}$ using only $16$ selected basis configurations, which is far higher in accuracy than any of the randomly selected basis set of the same size.
This roughly corresponds to the accuracy achieved with the randomly selected basis configurations giving the maximal marginal likelihood which are however an order of magnitude larger than the basis set selected by the RVM.
Selecting the basis configuration by application of the RVM, as an optimization of the marginal likelihood with respect to the variances of the prior distributions for the weights, thus makes it possible to learn a sparse and optimal GPS form from given target state data in a deterministic and efficient manner.

\subsection{Optimization of hyperparameters} \label{sec:Hyperparams}

\begin{figure*}
    \centering
    \includegraphics[width=\textwidth]{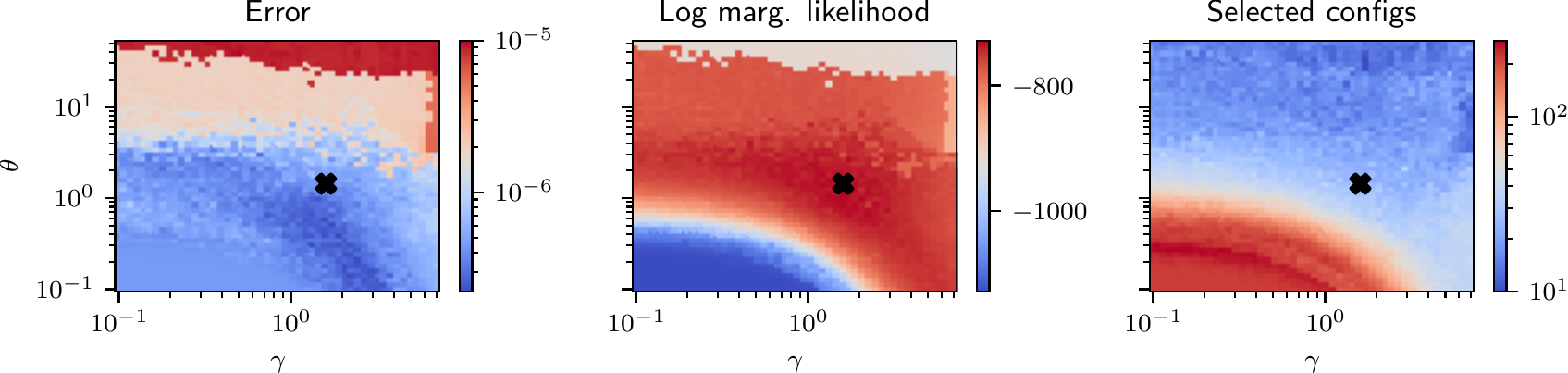}
    \caption{Heatmaps visualizing the mean squared error (left plot), the logarithm of the marginal likelihood (central plot) and the number of selected basis configurations (right plot) of a learned GPS with respect to kernel hyperparameters, $\theta$ and $\gamma$. The GPS is learned with the RVM on the exact ground state amplitudes of an eight site Hubbard chain at $U/t = 8$ using an accuracy parameter of $\tilde{\sigma}^2 = 10$. Larger values of the hyperparameters suppress the fit of more complex (high-rank and long-range) features, and reduces the number of basis configurations selected by the RVM. The position corresponding to the maximum of the log marginal likelihood is represented by a cross on all plots.}
    \label{fig:likelihood_vs_hyp_param}
\end{figure*}

With the type II maximum likelihood scheme resulting in models representing an optimal balance between sparsity and accuracy, we also aim to optimize the remaining dependence of the model, the hyperparameters, in a similar priciple of maximal marginal likelihood via the RVM.
Figure~\ref{fig:likelihood_vs_hyp_param} shows the model mean squared error with respect to the target state, the number of basis configurations selected and the final logarithm of the marginal likelihood obtained, as the two kernel hyperparameter are varied, and the RVM procedure used to select the basis configurations for each point. The hyperparameter controlling the weighting of the order or rank of the features ($\theta$) varies between $0.1$ and $50$, while the distance weighting hyperparameter ($\gamma$) is varied between $0.1$ and $7$.
As in the previous example, the target state from which the representation is learned with the RVM corresponds to a ground state of an eight site Hubbard chain in the strongly correlated regime at $U/t=8$ and we fix the variance parameter to $\tilde{\sigma}^2 = 10$.

The learned models reach mean squared errors on the target state ranging between $\approx 2 \times 10^{-7}$ and $\approx 10^{-5}$ with selected basis sets varying in size between $273$ and $10$.
For the chosen example, the highest accuracy is achieved in intermediate parameter regimes of the kernel hyperparameters.
Choosing $\theta \to 0$, $\gamma \to 0$, the values of the kernel function vanish for all but symmetrically related pairs of configurations.
As a consequence approaching this limit results in a large fraction of all potential basis configurations selected by the RVM.
In contrast, with increasing values of the correlation order weighting parameter $\theta$ (suppressing the fitting of higher rank correlations) as well as increasing values of the correlation range weighting parameters $\gamma$ (suppressing the fitting of longer ranged correlations), the model becomes less complex, requiring fewer configurations to be selected by the RVM and resulting in more compact representations.

Going to very large values of $\theta$ and $\gamma$ reduces the complexity of the underlying model by preferentially weighting the fit to simpler features. This decreases the accuracy of the final GPS with respect to the target state. However, there is a minimum in the error at a particular choice of hyperparameters, which also roughly corresponds to the maximum in the log marginal likelihood. This intermediate hyperparameter regime corresponds to a $\theta$ and $\gamma$ of between one and two, and represents the optimal balance between the accuracy and compactness of the resulting GPS state, as enforced by the prior distribution on the weights. For this system, this corresponds to a fit of around $25$ basis configurations and reaches a mean squared error of $\approx 3.4 \times 10^{-7}$, as denoted by the cross in Fig.~\ref{fig:likelihood_vs_hyp_param}.

\begin{figure}
    \centering
    \includegraphics[width=\columnwidth]{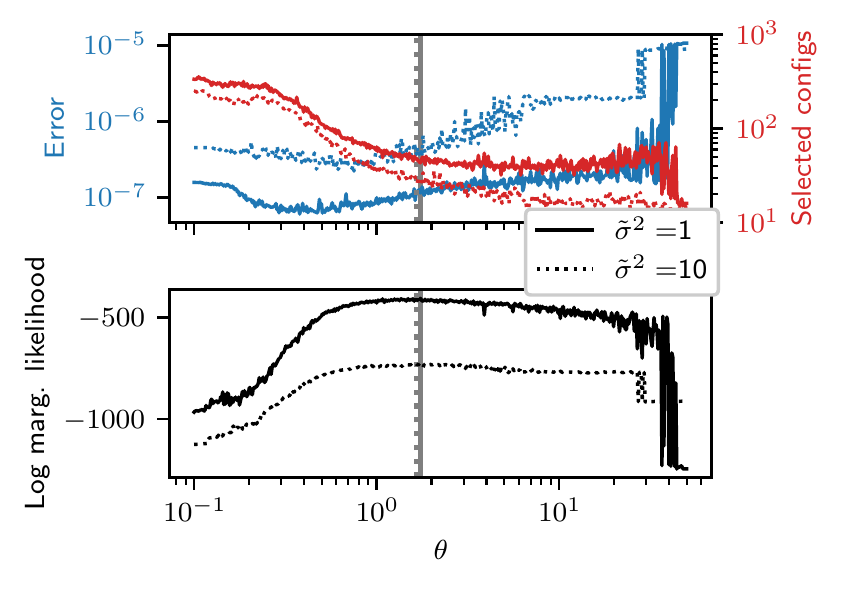}
    \caption{Number of selected configurations (top panel, red), mean squared error (top panel, blue) and marginal likelihood (bottom panel) against $\theta$ for the fit of the same target state as in Fig.~\ref{fig:likelihood_vs_hyp_param}, using a fixed $\gamma = 1$ and two different values $\tilde{\sigma}^2 = 1$ (solid lines) and $\tilde{\sigma}^2= 10$ (dotted lines). Vertical lines correspond to the value of $\theta$ where the marginal likelihood is maximal.}
    \label{fig:likelihood_sparsity_vs_bheta}
\end{figure}

More detailed insight into the relationship between fit accuracy, model complexity and marginal likelihood can be gained from the results of Fig.~\ref{fig:likelihood_sparsity_vs_bheta}. In this, these quantities are shown for the same target state, as the kernel complexity is varied via the $\theta$ hyperparameter for a fixed $\gamma = 1$. 
These quantities are also shown for two different choices of the accuracy parameter $\tilde{\sigma}^2 = 1$ and $\tilde{\sigma}^2 = 10$, which formally controls the desired variance of the likelihood of modelling the wave function amplitudes of the target function.
The smaller value of $\tilde{\sigma}^2$ leads to a more accurate but less sparse model across the whole displayed range of $\theta$, allowing $\tilde{\sigma}^2$ to effectively control the desired accuracy-sparsity trade-off.
In agreement with the results of Fig.~\ref{fig:likelihood_vs_hyp_param}, the number of basis configurations overall decreases with increasing values of $\theta$, as the model becomes simpler.
However for both choices of $\tilde{\sigma}^2$, the obtained mean squared error is minimal in a regime of small $\theta$ values between $\approx 0.3$ and $\approx 0.8$.

The fact that the error does not decrease further for more complex models is due to the modelling assumptions in the Bayesian learning, where a larger value of $\tilde{\sigma}^2$ (approximately) corresponds to a larger variance of the modelled likelihood of the data, i.e. the accuracy to which we attempt to reproduce the training data.
\toadd{The inferred weights for a given basis set do then not necessarily give the minimal least squared error with respect to the training data since essentially a Gaussian noise model is assumed for the fitted data which is governed by $\tilde{\sigma}^2$.}
Depending on the chosen noise parameter, the error of the model does therefore not automatically approach zero in the limit where (almost) all basis configurations are selected by the RVM, due to this finite noise model controlling the desired accuracy.
Nonetheless, for both choices of $\tilde{\sigma}^2$, the marginal likelihood takes the maximal value in an intermediate regime of $\theta$ where a good accuracy is achieved with a compact basis set.
The value of $\theta$ which is associated with the maximum value of the marginal likelihood is indicated by vertical lines in the figure.
For both displayed values of $\tilde{\sigma}^2$ the marginal likelihood is maximal in the regime of $\theta \approx 2$ resulting in a basis set of $27$ basis configurations with an error of $\approx 4 \times 10^{-7}$ for $\tilde{\sigma}^2 = 10$.
Choosing a smaller error parameter $\tilde{\sigma}^2 = 1$, the GPS model giving the maximal marginal likelihood is less sparse but also more accurate underlining the significance of the noise parameter $\tilde{\sigma}^2$ which can be used to tune the target accuracy which we want to achieve with the learned GPS obtained by the maximization of the marginal likelihood.

In order to obtain a final compressed GPS form of a given quantum state, we therefore also optimize the marginal likelihood with respect to the kernel hyperparameters in the application of the RVM.
The only input parameter for this algorithm is the parameter $\tilde{\sigma}^2$ implicitly controlling the target accuracy we aim to achieve with the fit.
We optimize the kernel hyperparameters by running the RVM multiple times using different hyperparameters and use the model giving the largest maximum likelihood as the final representation learned from the data.
In order to find the maximum of the marginal likelihood (or a good approximation thereof) with as few repeated applications of the RVM as possible
, we apply the \textit{sequential model based optimization} scheme (also known as Bayesian optimization) based on the \textit{Tree of Parzen Estimators} \cite{NIPS2011_4443, pmlr-v28-bergstra13} as implemented in the \textit{hyperopt} python package \cite{Bergstra_2015}.
In this scheme the space of possible hyperparameters is sampled according to an initial assumed distribution of the parameter space while also taking into account the evaluation history from previously sampled values of the kernel hyperparameters.
Similar to the introduced Bayesian learning approach, this is done by modelling underlying probability distributions for the kernel hyper parameters and obtained log marginal likelihood values based on the observations and chosen prior distributions for the hyperparameters.
New hyperparameter values are then sampled according to their associated expectation value of the improvement of the log marginal likelihood with respect to a chosen threshold value.
Based on the general scaling behaviour as it is shown in Fig.~\ref{fig:likelihood_vs_hyp_param}, we choose log-uniform priors for $\theta$ and $\gamma$.

\subsection{Full optimization of the GPS for correlated systems} \label{sec:FullOpt}

\begin{figure}
    \centering
    \includegraphics[width=\columnwidth]{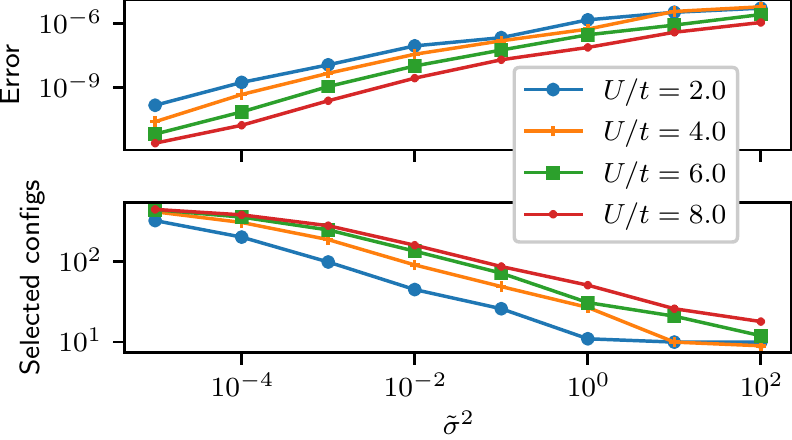}
    \caption{Mean squared error (top panel) and number of selected basis configurations (bottom panel) obtained for a GPS representation learnt from the ground state of an eight site half-filled Hubbard chain at $U/t=2, 4, 6$ and $8$. Horizontal axes corresponds to the chosen accuracy parameter $\tilde{\sigma}^2$ defining the noise model in the regularization of wave function amplitudes.}
    \label{fig:learning_curve}
\end{figure}

We can now apply Bayesian optimization of weights, basis set configurations, and kernel hyperparameters, to lattices at a variety of correlation strengths, to consider the accuracy and compactness of the resulting GPS representation.
Figure~\ref{fig:learning_curve} shows the mean squared error with respect to the exact target state, the number of selected basis configurations, as well as the optimized kernel hyperparameters, for different values of the variance $\tilde{\sigma}^2$.
The target state learned is the Hubbard model ground state of the eight site system at $U/t = 2$, $4$, $6$ and $8$.
As expected, for all values of $U$, the error decreases monotonically with decreasing values of $\tilde{\sigma}^2$, with the number of configurations required in the basis set correspondingly increasing.

For the most strongly correlated system at $U=8t$, we obtain a worst mean squared error of $\approx 10^{-6}$ with only $18$ basis configurations when $\tilde{\sigma}^2 = 100$. This error in the wave function prediction across all configurations can be systematically improved by almost $6$ orders of magnitude (down to an error of only $10^{-12}$ -- close to numerical precision) by decreasing the variance/noise down to $\tilde{\sigma}^2=10^{-5}$, at which point the GPS requires $446$ basis configurations. This is compared to the full complexity of the wave function, which for this small system with $8$ Fermionic degrees of freedom, requires $4,900$ configurational amplitudes (reducing to $618$ if only symmetry-inequivalent configurations are considered).
However, while this compression is not in itself particularly noteworthy, it should be stressed that due to the size extensive nature of the state, it is not anticipated that the size of the basis set for a desired accuracy will increase with the size of the system. This observation is key to the development of numerical algorithms based on the GPS construction in Ref.~\onlinecite{Glielmo2020}, well beyond the system sizes of conventional calculation.

Furthermore, while the specific accuracy and basis set sizes vary with $U/t$ in Fig.~\ref{fig:learning_curve}, these differences are relatively minor, and the overall trend is is largely independent of the interaction strength. This bodes well for the GPS within the Bayesian learning framework to be applicable to significantly different physical regimes with similar accuracy.
The optimized kernel hyperparameters are also largely independent of the specified noise $\tilde{\sigma}^2$, potentially rendering the optimization of the hyperparameters an optional step.
The optimized values of $\gamma$ range between $\approx 0.9$ and $\approx 2.8$, resulting in a suppression of long range features which is similar to the generic distance weighting proposed in Ref.~\onlinecite{Glielmo2020} of $1/|r_i - r_0|$. Of course, further flexibility can also be included into the kernel function with additional hyperparameters, such as a fully flexible distance weighting \cite{Glielmo2020}. The scope to improve the GPS form with more general `kernel learning' of the model complexity will be explored in future work.

\section{Extraction of physical information from a Gaussian Process State}

Having defined a general scheme to learn a compact GPS representation including its kernel hyperparameters from given wave function data, we then in theory have access to the wave function, and therefore all physical (static) observables.
However, by learning the kernel hyperparameters, we also potentially have access to a characterization of the state which does not correspond simply to a physical observable, but can still provide insight into the system.
For the hyperparameters used in this work, the value of $\theta$ can provide information on the order or rank of correlations, as well as $\gamma$ providing a physical length scale for these correlations required to reach the target accuracy with the model. 
The optimized hyperparameters therefore provide meaningful insight about the underlying physics of the modelled target state.

\begin{figure}
    \centering
    \includegraphics[width=\columnwidth]{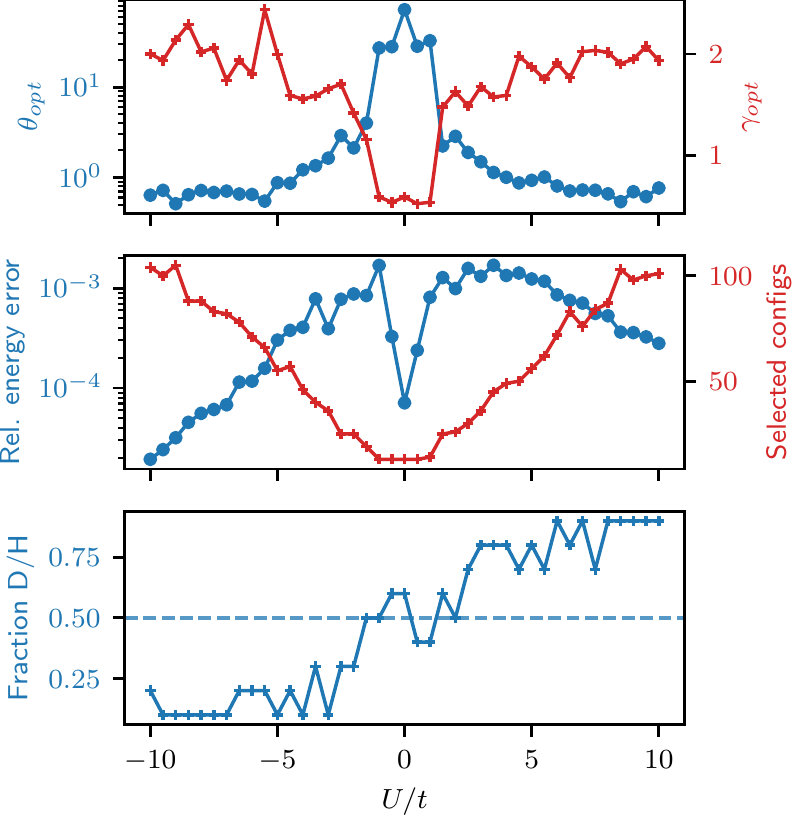}
    \caption{Optimized hyperparameter values $\theta_{opt}$ (top panel, blue) and $\gamma_{opt}$ (top panel, red), relative energy error (center panel, blue) and number of selected basis configurations (center panel, red) as well as the fraction of selected basis configurations which have a doublon or holon occupancy on the reference site (bottom panel) for learned GPS representations of the eight site Hubbard chain ground state at different effective interaction strengths of $U/t$. The fraction of basis configurations with a doublon or holon occupancy of the reference site is computed with respect to the ten most relevant configurations, as given by the smallest optimized values $\alpha_{\mathbf{x}'}$. The accuracy value for learning the GPS is chosen as $\tilde{\sigma}^2 = 0.1$.}
    \label{fig:obtained_hyperparameters}
\end{figure}

The top panel of Fig.~\ref{fig:obtained_hyperparameters} shows the trends in the optimized hyperparameters obtained from learning the ground state of the Hubbard chain for different values of the interaction strength $U/t$.
The optimized $\gamma$ defines an (inverse) lengthscale for the correlations, and shows an overall increase across the range of positive $U/t$ going up from a value of $\approx 0.6$ at $U/t=0$ to a value of $\approx 1.9$ at $U/t=10$. In contrast, the optimal $\theta$ hyperparameter, defining an (inverse) correlation order weighting, decreases as the correlation strength increases, from $\theta \approx 72$ at $U/t = 0$ to $\theta \approx 0.8$ at $U/t = 10$. 
The optimized kernel parameter therefore reflects the change in correlations encoded in the wave function character, transitioning from a low-order long-range correlation characteristic of the mean field limit where kinetic energy considerations dominate, to high-order but short-ranged correlation properties in the regime of strong interaction where strong but short-ranged magnetic and charge quantum fluctuations dominate.

In the centre panel of Fig.~\ref{fig:obtained_hyperparameters} we report the relative error in the energy of the modelled state (rather than the wave function amplitude error as shown previously), as well as the number of selected basis configurations obtained in the GPS representations, as the correlation strength is changed.
In the repulsive regime, the size of the basis set increases largely monotonically with interaction strength, ranging from $13$ configurations at $U/t=0$ to a total of $101$ selected configurations at $U/t = 10$.
While this increase of the basis set size is also reflected by a decrease of the relative energy error in the strongly-correlated regime of $U/t \gtrapprox 3.5$, this is not true for smaller values.
The maximum of the relative energy error at around $U/t \approx 3.5$ points to the significant challenge of modeling wave functions in intermediately correlated regimes with a highly sparse GPS, where there is competition between long-range and short-range correlations. While it is not surprising that this intermediate regime, where no single physical feature dominates, is one where the error in the GPS is highest, this may also be able to be addressed in the future by finding alternative kernel parameterizations which are more flexible and better suited for a description of this physics.

Insight about specific correlation patterns, beyond just the values of the hyperparameters, can also be extracted from the unique GPS model, via an analysis of the basis set selected by the RVM with the optimized hyperparameters.
To visualize the space of all many-body configurations of the underlying Hilbert space, we find a two-dimensional representation based on the \textit{t-distributed stochastic neighbor embedding} (t-SNE) \cite{Maaten2008} with the chosen kernel hyperparameters.
In order to apply the t-SNE, we divide the Hilbert space into four different classes of configurations based on the occupancy of the central reference site used to define the (non-symmetrized) 
kernel of Eq.~\eqref{eq:final_kernel}.
The t-SNE algorithm then maps all configurations of each class into a two-dimensional space according to the scaled Hamming distance metric, $h(\mathbf{x}, \mathbf{x}')/\theta$, used in the kernel function.
If two configurations have a small scaled Hamming distance, resulting from a large overlap in the feature space, then these two configurations are also likely mapped to geometrically close points in the t-SNE representation.

\begin{figure}
    \centering
    \includegraphics[width=\columnwidth]{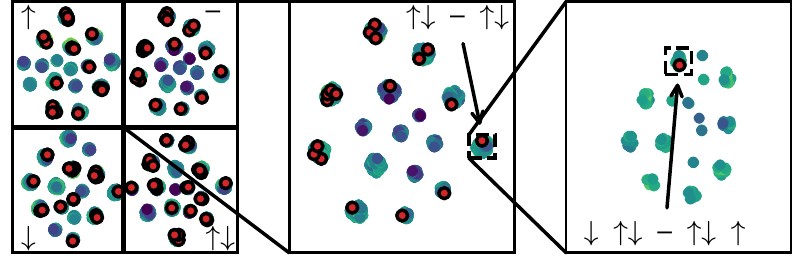}
    \caption{Two-dimensional t-SNE representation of the many-body configurations of an eight site Hubbard model, where the panels show increasing resolution of individual parts of the Hilbert space. The Hilbert space is initially separated into four different classes of configurations based on the occupancy of the reference site as shown, which is further divided hierarchically into 16 further sub-clusters of configurations. The distance metric for the t-SNE of each class is chosen according to the scaled Hamming distance used in the kernel with hyperparameters chosen by learning the ground state at $U/t= 8$ with $\tilde{\sigma}^2=0.1$. The configurations selected by the RVM are marked in red and the color coding of the scatterpoints is chosen according to the magnitude of the associated probability amplitude of the configurations for this system. Features around the reference site defining each cluster of configurations are also indicated in each panel.}
    \label{fig:clustering_of_configs}
\end{figure}

The t-SNE distributions for the Hilbert space basis configurations of the eight site Hubbard chain used in this work at $U/t=8$ is shown in Fig.~\ref{fig:clustering_of_configs}.
Each scatter point represents one many-body configuration mapped to a position on the visualized 2D plane by the t-SNE algorithm.
As can be seen, the form of the kernel results in a hierarchical clustering of the configurations, with a self-similar structure of the clusters on different levels. Each cluster of configurations corresponds to a further $16$ clusters when zoomed in to the next level. This form arises due to the 16 different additional correlation features which arise when considering all possible occupancies of the site one further from the reference ($16=4^2$ for all Fock states of the two additional sites at the next distance from the reference site). This structure arises mathematically from the distance weighting of the features describing the scaled distances as $\vert d \vert^{-\gamma}$, weighting non-matching occupancies less for sites further away from the reference central site.
Consequently, this results in a fractal structure where each cluster can again be separated into different clusters based on the occupancies of the sites directly adjacent to the sites characterizing the higher level cluster, as shown in Fig.~\ref{fig:clustering_of_configs}.

\begin{figure*}
    \centering
    \includegraphics{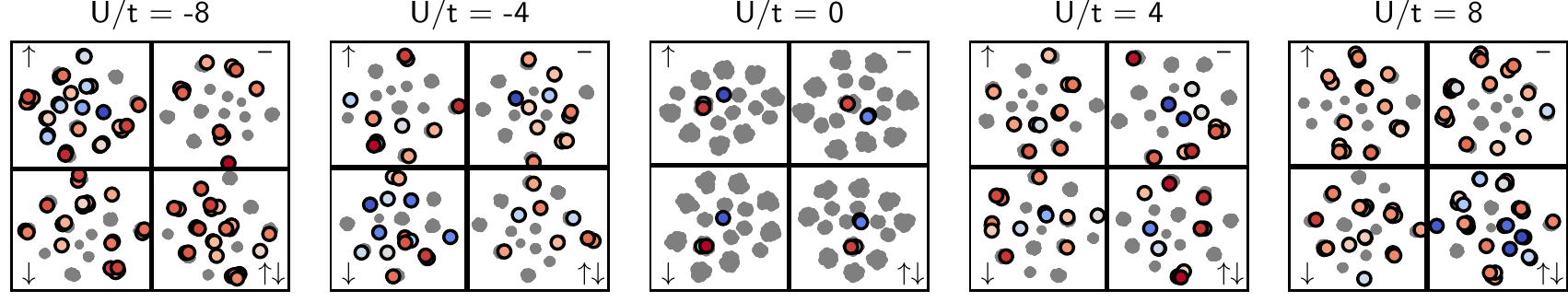}
    \caption{Selected configurations (colored) in the two-dimensional space of the t-SNE representation of all Hilbert space configurations (grey) for different values of $U/t$. The selected configurations as well as the underlying kernel hyperparameters are obtained by training the GPS on the ground state of the eight site Hubbard model with $\tilde{\sigma}^2= 0.1$. The color of each point for the selected basis configuration indicates the value of the associated weight, with the warmer colors corresponding to larger weights and the colorscale rescaled for each individual value of $U/t$.}
    \label{fig:rvm_positionining_clusters}
\end{figure*}

While this pictorial representation of the kernel function between configurations is interesting in its own right, we are also then able to use this representation to analyze the position of the specific basis configurations selected by the RVM to characterize the GPS. These configurations will likely exhibit the most relevant correlation features present in the state. In Fig.~\ref{fig:clustering_of_configs}, these are marked in red for the $U/t=8$ GPS. However, of more interest is how this selection of the basis configurations changes with the character of the wave function and correlation strength. This is shown in Fig.~\ref{fig:rvm_positionining_clusters} for representations learned from ground states at different values of the interaction strength $U/t$.

As well as indicating the selected basis configurations in Fig.~\ref{fig:rvm_positionining_clusters}, the choice of colour reflects the magnitude of the optimized weights associated with these basis configurations. The largest positive weights are denoted with warmer (red) colours, while the largest negative weights are denoted with colder (blue) colours, while unselected basis configurations are shown in grey. What is found, is that in the {\em attractive} Hubbard model, characterized by $U/t < 0$, the selected basis configurations predominantly reside in the sectors of the Hilbert space corresponding to singly occupied reference sites. As $U/t$ increases and the system transitions into the more common {\em repulsive} Hubbard regime, the chosen basis configurations with dominant weights are rather found in the sectors of the Hilbert space with unoccupied or doubly occupied central sites. This indicates a change in dominant character of the GPS whereby it goes from trying to suppress correlation features with singly occupancy in the attractive case (thereby promoting pairing order of the attractive electrons), to suppression of double occupancy/holon character (thereby promoting single occupancy and magnetic order) in the repulsive Hubbard model.


The observation that with increasing values of $U/t$ predominantly basis configurations with empty or doubly occupied reference sites are selected can be quantified by analyzing the fraction of selected basis configurations with such reference site occupancies.
This is shown in the bottom panel of Fig.~\ref{fig:obtained_hyperparameters}, and confirms that in the strongly attractive regime with $U/t \lessapprox -4$, approximately $10 - 20 \%$ of the $10$ most relevant selected basis configurations (i.e. configurations with the smallest optimized $\alpha$ values) have a doublon or holon occupancy on the reference site.
For increasingly repulsive effective interaction strengths, the fraction of basis configurations with such doublon or holon occupancies among the most relevant basis configurations increases, resulting in a fraction of $\approx 90 \%$ for $U/t \gtrapprox 8$, confirming the hypothesis from Fig.~\ref{fig:rvm_positionining_clusters}. Further insight into longer-ranged orderings, including striped or inhomogeneous character in the dominant correlation features can be extracted by considering the prevalence and weighting of longer range features present in the chosen basis configurations.
This analysis of the learned GPS underlines how the physically motivated definition of the model makes it possible to easily interpret the character of the correlations that emerge from this Bayesian learning scheme, and to gain physical insight into the target state beyond just the extraction of accurate physical observables.

\section{Conclusion}
In this work, we have given an overview of how a supervised learning scheme can be applied to learn a compact and accurate representation of correlated many-body states in the form of a Gaussian Process State from given data of the target state.
Due to its particular form, corresponding to a linear model in some higher dimensional feature space, the GPS is particularly well suited for regression with statistical methods based on Bayesian inference.
A central conclusion of this work is that the optimization of the marginal likelihood for basis selection yields representations balancing the sparsity and accuracy of the model without the need of additional cross validation of the model.
The introduced supervised learning scheme based on the relevance vector machine therefore provides an efficient tool to compress given wave function data to a compact but highly expressive functional model in the form of a GPS.
Because the central building block of the GPS, the kernel function, can be motivated from physical insight into the feature space it implicitly represents, it is possible to easily extract meaningful characteristics of the modelled state from the learned representation and obtain additional insight of the underlying physical properties of the state.
In contrast to ansatzes based on explicit physical representations such as the Jastrow ansatz, the GPS is however not limited or biased in its expressibility making it possible to model arbitrary correlation properties in the state with the chosen kernel function, which can however still be evaluated in polynomial time.

The compression scheme presented in this work can easily be applied to a range of different applications where it is desired to obtain a compact functional form of a wave function from some limited observed data.
In addition to the previously developed numerical schemes within ground states of larger lattice systems which do not allow for an exact numerical treatment, the compact and interpretable functional form of the GPS could similarly be used for problems such as finding higher excited states, modeling many-body dynamics \cite{Hallberg1995, Hendry2019, Zhang2020}, describing ab-initio chemical systems \cite{Friesner2005, Schuett2019, Choo2020, Pfau2019, Hermann2019} or using them within quantum state tomography \cite{Vogel1989, Cramer2010, Torlai2018}.
Although the range of potential application is similar as for the NQS, the particular form of the GPS model could make it a favourable and more interpretable ansatz in some instances so that the GPS complements the set of machine learning inspired many-body representations.

\begin{acknowledgements}
The authors thank G\'{a}bor Cs\'{a}nyi for useful discussions in the development of this work.
G.H.B. gratefully acknowledges support from the Royal Society via a University Research Fellowship, and funding from the Air Force Office of Scientific Research via grant number FA9550-18-1-0515.
The project has also received support from the European Union's Horizon 2020 research and innovation programme under grant agreement No. 759063.
We are grateful to the UK Materials and Molecular Modelling Hub for computational resources, which is partially funded by EPSRC (EP/P020194/1).
\end{acknowledgements}

%

\end{document}